\documentclass[aps,
onecolumn,groupedaddress,nofootinbib,nobalancelastpage,nobibnotes]{revtex4}
\pdfoutput=1
\usepackage{amsmath,amsfonts,amssymb,mathrsfs,graphicx,color}
\usepackage[squaren]{SIunits}
\usepackage{verbatim}
\usepackage{enumerate}
\usepackage{graphicx}
\usepackage[hyperfootnotes=false]{hyperref}
\usepackage{xcolor}
\usepackage{slashed}
\usepackage[utf8]{inputenc}

\newcommand{\ie}{{\it i.e.}}

\newcommand{\eg}{{\it e.g.}}

\newcommand{\etc}{{\it etc.}}
\newcommand{\eq}{Eq.}

\newcommand{\fig}{Fig.}

\newcommand{\Ref}{Ref.}
\newcommand{\Refs}{Refs.}

\newcommand{\equ}[1]{\eq~(\ref{equ:#1})}
\newcommand{\figu}[1]{\fig~\ref{fig:#1}}

\newcommand{\bi}{\begin{itemize}}
\newcommand{\ei}{\end{itemize}}

\newcommand{\revise}[1]{#1} 

\begin{document}


\title{Nuclear Physics Meets the Sources of the Ultra-High Energy Cosmic Rays}

\author{Denise Boncioli}
\affiliation{DESY, Platanenallee 6, D-15738 Zeuthen, Germany}

\author{Anatoli Fedynitch}
\affiliation{DESY, Platanenallee 6, D-15738 Zeuthen, Germany}

\author{Walter Winter\footnote{Correspondence should be addressed to denise.boncioli@desy.de (Denise Boncioli), anatoli.fedynitch@desy.de (Anatoli Fedynitch), walter.winter@desy.de (Walter Winter)}}
\affiliation{DESY, Platanenallee 6, D-15738 Zeuthen, Germany}

\date{\today}

\begin{abstract} 
The determination of the injection composition of cosmic ray nuclei within astrophysical sources requires sufficiently accurate descriptions of the source physics and the propagation -- apart from controlling astrophysical uncertainties.
We therefore study the implications of nuclear data and models for cosmic ray astrophysics, which involves the photo-disintegration of nuclei up to iron in astrophysical environments. 
We demonstrate that the impact of nuclear model uncertainties is potentially larger in environments with non-thermal radiation fields than in the cosmic microwave background. We also study the impact of nuclear models on the nuclear cascade in a gamma-ray burst radiation field, simulated at a level of complexity comparable to the most precise cosmic ray propagation code.
We conclude with an isotope chart describing which information is in principle necessary to describe nuclear interactions in cosmic ray sources and propagation.
\end{abstract}

\maketitle

\section{Introduction}
Particles from space reaching the Earth with energies higher than $10^{9}$ GeV are detected by ultra-high energy cosmic ray (UHECR) observatories such as the Pierre Auger Observatory~\cite{ThePierreAuger:2015rma} and the Telescope Array (TA) experiment~\cite{AbuZayyad:2012kk}. UHECRs are expected to be accelerated in astrophysical sources and to travel through extragalactic space before they hit the Earth's atmosphere; they can interact with photons in both environments. The primary composition of UHECRs is still unknown; however, the mass composition measured by the Auger Observatory indicates heavier elements at the highest energies beyond $10^{9.3}$~GeV~\cite{Aab:2014kda,Aab:2014aea,Aab:2016htd,Porcelli:2015}, \ie, significantly heavier than helium and at most as heavy as iron. The study of interactions of nuclei is therefore critical for our understanding of cosmic ray astrophysics both within sources and during propagation.

Most of the literature, as for example \cite{Allard:2008gj,Taylor:2011ta,Taylor:2013gga,Taylor:2015rla,Fang:2013cba,Aloisio:2013hya,diMatteo:2015,Peixoto:2015ava,Aab:2016zth}, focuses on finding the right cosmic ray composition injected from the sources into the intergalactic medium, propagating it through  the cosmic microwave background (CMB) and the extragalactic background light (EBL), which are thermal target photon fields, \ie, relatively strongly peaked. The long-term vision is, however, to trace back the cosmic ray composition into the \revise{sources}, which requires a combined source--propagation model; see \eg\ \Refs~\cite{Globus:2014fka,Unger:2015laa}. Such models face several challenges, including 1) largely unknown astrophysical environments and uncertainties, 2) limited computational resources for detailed simulations and parameter space studies, and 3) a complicated interplay of radiation processes -- especially in the source, where the photon fields are often non-thermal (power laws) and the physical processes have been less studied.  One of the main challenges is therefore to walk the line between precision and efficiency to overcome problems 2) and 3), while new insights on the astrophysical parameters 1) are to be obtained from multi-messenger observations. An example is \Ref~\cite{Baerwald:2014zga}, where, for a pure proton composition, constraints on the astrophysical parameters are derived from cosmic ray and neutrino observations. \revise{In order to extend such approaches to heavier compositions}, the physical processes have to be controlled with a precision as high as possible, where the photo-disintegration of nuclei plays the leading role. While the required target precision is arguable, a first bottleneck  is the description of the source as accurate as the propagation from source to detector -- where the target photon environment can be very different, and which has been well studied.

In this work, we focus on the photo-disintegration of nuclei, which has been extensively studied in the CMB and EBL, where it is the dominant process changing the mass composition of the nuclei.
The leading contribution to photo-disintegration is an excitation called ``giant dipole resonance'' (GDR)~\cite{Goldhaber:1948zza}, which can be interpreted as a vibration of the bulk of protons and neutrons leading to a resonant structure. This process occurs above $\sim$8~MeV (energy in the nucleus' rest frame) and causes the disruption of the primary nucleus with the emission of one or two nucleons. At higher energies the ``quasi-deuteron'' (QD) process dominates, where the photon interacts with a nucleon pair followed by consequent ejection of nucleons or light fragments. Note that we do not consider astrophysical situations where disintegration is dominated by even higher energy processes, such as baryonic resonances, at energies beyond 150~MeV.

A frequently used model in cosmic ray astrophysics is Puget-Stecker-Bredekamp (PSB)~\cite{Puget:1976nz}, that relies on choosing one isotope for each mass number $A$, and a unique disintegration chain populated through subsequent emission of nucleons. This approach is implemented for cosmic ray propagation in the {\it SimProp} software~\cite{Aloisio:2012wj}. A more sophisticated approach, based on the TALYS nuclear reaction program~\cite{Koning:2007}, 
is implemented in the cosmic ray propagation software CRPropa2 and 3~\cite{Kampert:2012fi,Batista:2016yrx}, which includes 183 isotopes and 2200 channels for the photo-disintegration. Differences in modeling the interactions affect the observables (as energy spectrum and composition), and consequently have an impact on the interpretation of UHECR measurements~\cite{Aab:2016zth,diMatteo:2015,Batista:2015mea}. 

UHECRs are expected to be accelerated in astrophysical sources, such as Gamma-Ray Bursts \revise{(GRBs; see \Ref~\cite{Meszaros:2006rc} for a review), Active Galactic Nuclei (AGNs), starburst galaxies, or jets produced in other cataclysmic events (mergers of neutron stars or black holes, tidal disruptions of massive stars getting too close to a super-massive black hole, \etc) -- to name a few examples. In sources such as GRBs or AGNs, they will disintegrate in the strong photon field present in the jets.} Examples for disintegration treatments \revise{in the sources} include~\cite{Globus:2014fka,Anchordoqui:2007tn}, where the GDR modeling follows \cite{Khan:2004nd} with a modified Gaussian parametrization of cross sections from \cite{Puget:1976nz}. The GDR resonance is even more simplified in \cite{Murase:2008mr,Murase:2010gj,Bustamante:2014oka} as a box function. Other authors use semi-analytical implementations of existing UHECR propagation codes (such as CRPropa)~\cite{Unger:2015laa}.
So far, however, the astrophysical sources have not been simulated with a complexity comparable to CRPropa for cosmic ray propagation -- including several hundred isotopes and the ten thousands of disintegration channels among them. That can be attributed to the fact that the target photon spectrum, relevant for the photo-disintegration, is {\em a priori} arbitrary, \ie, it can have a very different shape compared to that of the CMB. 

In this work, we present a description of the processes in the sources with a level of complexity of the interactions comparable to that of the most sophisticated cosmic ray propagation models. We review the available nuclear data necessary to construct and verify reliable interaction models, and we point out what information is missing from nuclear physics.

\begin{figure*}[t!]
\begin{center}
 \includegraphics[width=0.7\textwidth]{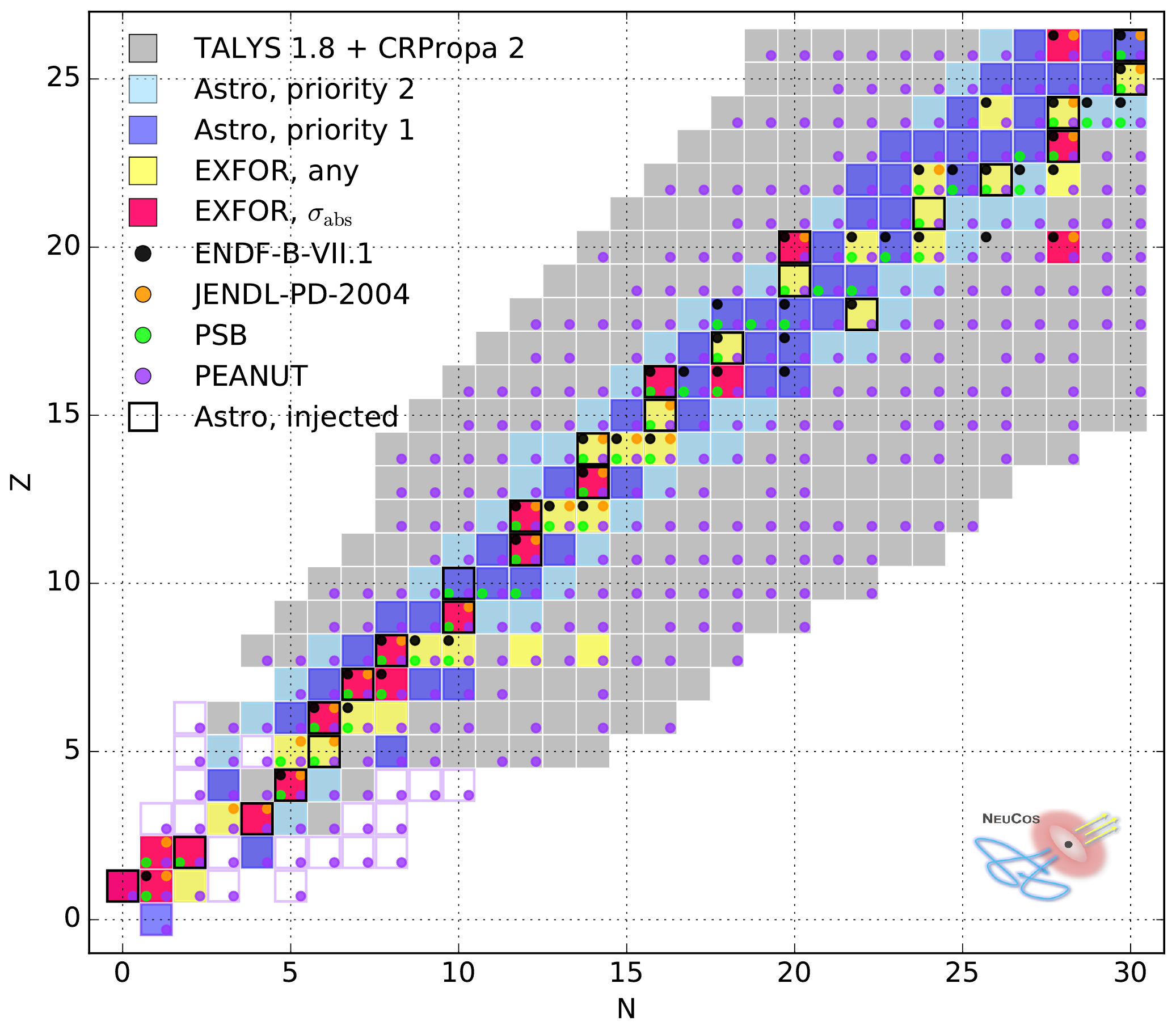} 
\end{center}
\caption{\label{fig:exfor_chart} Experimental situation versus astrophysical requirements for nuclear isotopes interesting for cosmic ray astrophysics (gray boxes, from TALYS \cite{Koning:2007} and CRPropa2 \cite{Kampert:2012fi}). Experimental measurements (from EXFOR database \cite{Otuka2014272}) are marked by red (yellow)  boxes if the total absorption (any inclusive cross section) has been measured. Theoretical models are marked by dots (ENDF \cite{Chadwick20112887}, JENDL \cite{JENDL-PD-2004}, PEANUT \cite{FLUKA_pd1,FLUKA_pd2}, PSB \cite{Puget:1976nz}). Calculations in cosmic ray astrophysics require the total and inclusive cross sections for the blue isotopes. 
 \revise{These isotopes have been obtained by recursively following all possible paths from all possible injection elements with different threshold multiplicities (for priorities~1 and~2), see main text for details.}
Injected isotopes are framed by black rectangles (\revise{we inject the most abundant stable isotope for each $Z$}). \revise{Violet framed rectangles} refer to very unstable isotopes integrated out in the disintegration chain.}
\end{figure*}

\section{Results}
\subsection{Situation on Experimental Data and Theoretical Models}

The experimental situation on photo-nuclear cross section data is shown in \figu{exfor_chart}, based on data from the most complete EXFOR database~\cite{Otuka2014272} (for details see the Supplementary Material). Of particular interest are measurements of the following cross sections in the GDR and QD energy bands ($E_\gamma < 150$ MeV): the total photo-absorption cross sections $\sigma_{\rm abs}$ (red boxes), the residual nucleus cross sections, such as \revise{for example} $\sigma_{\,^{27}\text{Al}}(\gamma,X + {^{24}\text{Na}})$ (where $X$ refers to any combination secondaries), and the inclusive light fragment cross sections, such as the neutron yield $\sigma_{\,^{27}\text{Al}}(\gamma,x \cdot \text{n}$)). In the following we refer to residual and light fragment cross sections as {\em inclusive} cross section $\sigma^\text{incl}_{^{27}\text{Al} \to  ^{24}\text{Na}}$ or $\sigma^\text{incl}_{^{27}\text{Al} \to  n}$, respectively  (yellow boxes, if at least one of these measured).
Data are sparse, and mostly available for stable elements along the main diagonal. Note that we did not find any $\sigma_{\rm abs}$ measurement for nuclides in the same isobar, \ie, two elements with the same mass number $A$. Furthermore,
note that in astrophysical environments, unstable isotopes gain importance, since all kinds of secondary nuclei are created in the disintegration chain and their lifetime is dilated by the relativistic boost. Therefore, these radioactive nuclei can re-interact with the photon field and create secondaries within the lifetime of the system. 

We also show the availability of nuclear models and data files in \figu{exfor_chart}, that use interpolated or fitted $\sigma_{\rm abs}$ where measurements are available. 
Unmeasured $\sigma_{\rm abs}$ are obtained from model evaluations of photo-neutron cross sections where available, otherwise from empirical parameterizations \cite{FLUKA_pd1}, implying that, in the absence of data, the cross sections further off the main diagonal are uncertain. Inclusive reaction cross sections are calculated with numerical or Monte Carlo 
codes, which are partially fine-tuned to data on branchings. \revise{We refer to EXFOR's $\sigma_{\rm abs}$ datasets as data, where real measurements are available (red boxes in \figu{exfor_chart}). Model evaluations for a subset of isotopes (yellow boxes) might exist and partially included in the cross section library of PEANUT. The latter subset demonstrates the potential for including corrections to the estimated $\sigma_{\rm abs}$ for isotopes, which are not covered by data.}

\begin{figure*}
\begin{center}
 \includegraphics[width=0.9\textwidth]{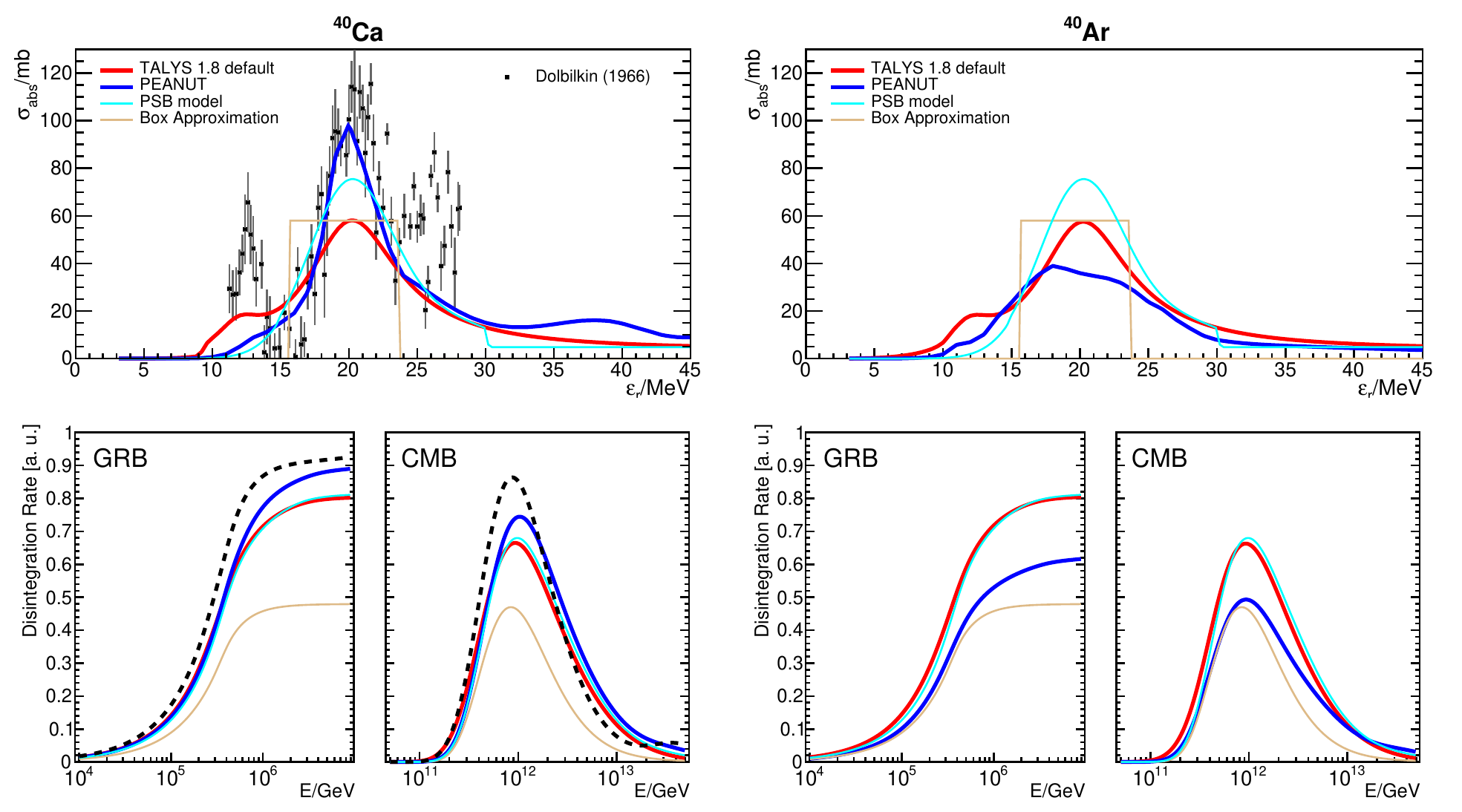} 
\end{center}
\caption{\label{fig:dis40} Comparison of cross sections (upper row) and disintegration lengths (lower row) for the isobar nuclides $^{40}$Ca (left column) and $^{40}$Ar (right column). The total absorption cross sections for photo-disintegration are shown as a function of the energy $\varepsilon_r$ in the nucleus' rest frame, where data are shown if available. \revise{The corresponding Lorentz factor of the nucleus is given by $\varepsilon_r/\varepsilon$, where $\varepsilon$ is the energy of the photons in the observer's frame (see the Supplementary Material for additional information)}. The different curves correspond to models as given in the plot legend, where the  GDR box approximation is based in the assumptions in \Ref~\cite{Murase:2010gj}.
\revise{The corresponding disintegration rates are calculated at redshift $z=0$ as a function of the observed energy; the corresponding Lorentz factor of the nucleus is given by $E/m_A$, where $m_A$ is the mass of the nucleus}. The disintegration rates are calculated for two different target photon spectra: for the GRB spectrum, a broken power law with spectral indices $-1$ and $-2$ and a break at 1~keV (energies in shock rest frame) has been assumed, whereas the CMB spectrum refers to the cosmic microwave background at redshift zero, \ie, a thermal target photon spectrum with $T=2.73$~K. Dashed lines refer to disintegration rates calculated for measured cross sections.}
\end{figure*}

\subsection{Cross sections and photo-nuclear disintegration rates} 
\begin{figure}
\begin{center}
\includegraphics[width=0.9\columnwidth]{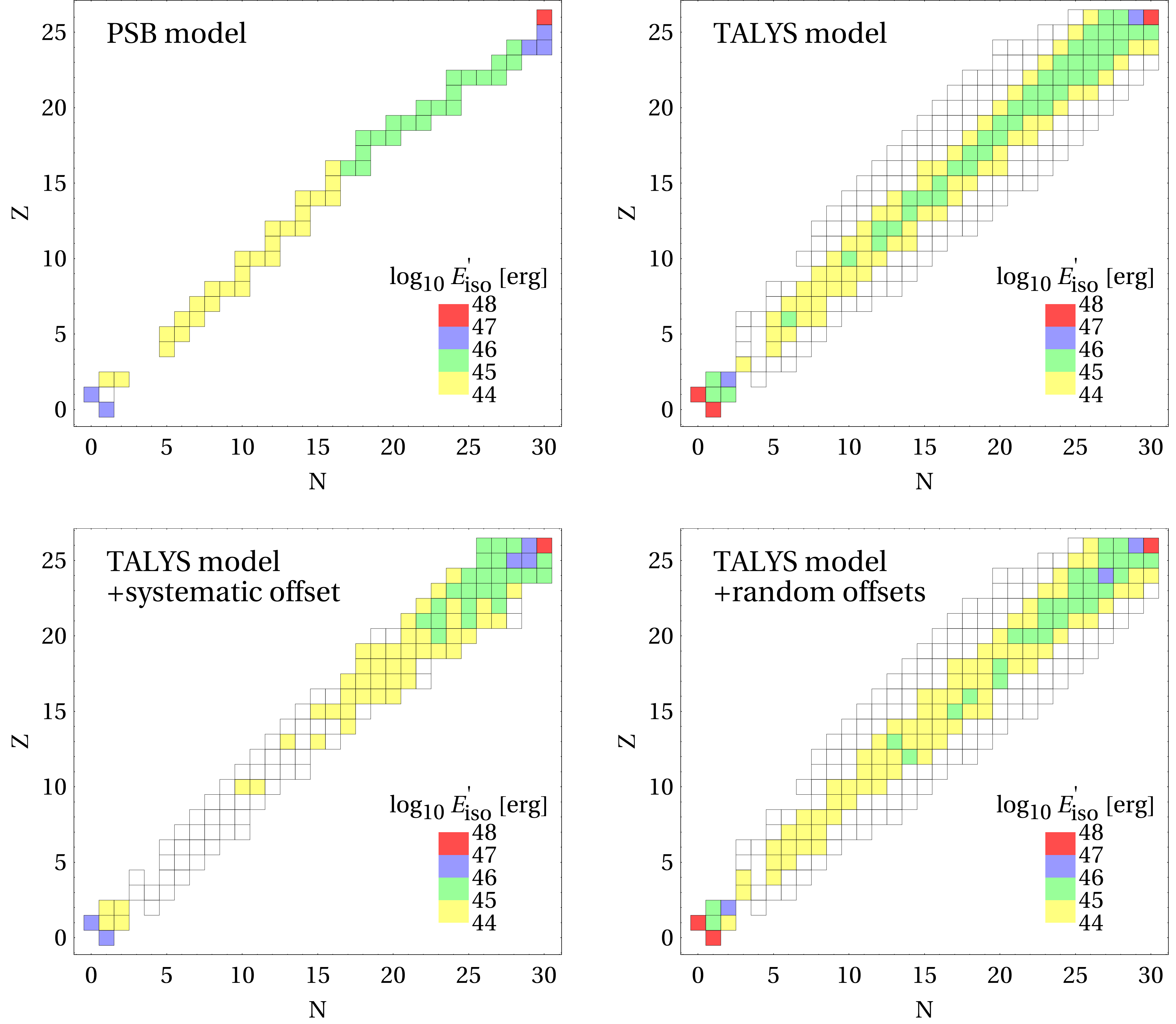} 
\end{center}
\caption{\label{fig:nuclei} Disintegration of $^{56}$Fe in a shell of a GRB, where the population of isotopes (in terms of the total energy per isotope and collision in the shock rest frame) is shown. \revise{The colors refer to the energy in the range indicated in the legend (between lower and upper shown value), and white boxes refer to energies below $10^{44} \, \mathrm{erg}$.} The upper panels correspond to the PSB \cite{Puget:1976nz} (left) and TALYS~\cite{Koning:2007} (right) (and CRPropa2~\cite{Kampert:2012fi} for the lighter elements) models. The lower panels correspond to the TALYS model for different systematics assumptions (see text for details). The GRB parameters are $L_{\gamma,\mathrm{iso}}=10^{52} \, \mathrm{erg/s}$, $\Gamma=300$, $t_v=0.01 \, \mathrm{s}$, $z=2$, and $E_{\mathrm{Fe,max}} \simeq 10^{11} \, \mathrm{GeV}$ (in the observer's frame) \revise{for an $E^{-2}$ injection spectrum}.}
\end{figure}

\revise{In the upper panels of \figu{dis40}, we illustrate one example of a typical situation on cross sections and their model representations for} two isobars with $A=40$: ${}^{40}$Ca is a double magic nucleus for which one photo-absorption cross section measurement is available \cite{Dolbilkin:1966}, while ${}^{40}$Ar is expected to have different properties due to a different shell structure. \figu{dis40} demonstrates that the TALYS predictions are almost independent of the isotope, while PEANUT, \revise{which is the base model for hadron-nucleus and photon-nucleus interactions in the FLUKA code \cite{Ferrari:2005zk,BOHLEN2014211}}, shows substantial changes between ${}^{40}$Ca to ${}^{40}$Ar and reproduces the data for ${}^{40}$Ca. \revise{The low-energy and high energy peaks observed in data are not present in the listed models, as well as in an evaluated dataset contained in EXFOR \cite{erokhova2003giant4421225}}. The PSB cross section is, by definition, the same for isobar nuclei. \revise{We estimate} uncertainty among different models to be of order two. An alternative case for mirror nuclei with $A=23$, where one would expect equal cross sections but finds differences in models is shown in the Supplementary Material. An equivalent comparison for $A=56$, that is frequently used in astrophysical calculations, is not possible due to absence of measurements. \revise{On the model side, TALYS and PEANUT predict a similar $\sigma_{\rm abs}$ and $\sigma_{\,^{56}\text{Fe}}(\gamma,1 \text{n})$, but differ by factor 2 in $\sigma_{\,^{56}\text{Fe}}(\gamma,1 \text{p})$ and $\sigma_{\,^{56}\text{Fe}}(\gamma,2 \text{n})$ for the standard parameter settings in TALYS. The PSB model, which has a one-dimensional disintegration chain in $A$ (see upper left panel of \figu{nuclei}), requires the emission of a proton to reach the next $A-1$ isotope $^{55}\text{Mn}$. This requires the model to overestimate the $\sigma_{\,^{56}\text{Fe}}(\gamma,1 \text{p})$ by factor 5 to 10, when compared to the more complete models.}

Here we illustrate the consequences of the experimental data availability and the prediction power of the models at the example of the photo-disintegration rate $\Gamma \sim c \, n_\gamma \, \sigma$ (see the Supplementary Material for details). 
The disintegration rate $\Gamma$ is indicative for the optical thickness for cosmic ray escape from the source $\tau \equiv R \cdot \Gamma/c$ (where $R$ is the source size), \ie,  $\tau\lesssim 1$ is a necessary condition for efficient cosmic ray escape. They are shown for two astrophysical environments, CMB and GRB, in the lower panels of \figu{dis40} -- corresponding to the cross sections in the upper panels. These examples are representative for a thermal (CMB) and a non-thermal (GRB) target photon field; we obtain similar results (to GRBs) for other non-thermal cases, such as active galactic nuclei. The uncertainties in the cross sections translate into similar uncertainties of the disintegration rates, which are sensitive to both the width and the threshold of the cross section peak(s). However, the effect from multiple peaks smears out due to pitch angle averaging, related to the (assumed) isotropic target photon distribution. Note that the relative impact of the models is qualitatively similar for the different target photon fields, but the quantitative impact can be larger for environments different from the CMB, which are more sensitive to the high-energy part of the cross section -- see GRB example.

Astrophysical applications involving sources of UHECR require predictability of the cross sections across a wider range of isotopes than in scenarios for which the available models were initially designed for. We find differences for individual isotopes, for example, total absorption cross sections seem consistently alike for different isobars in TALYS (unless tuned to a measurement), whereas differences are in some cases expected and predicted in PEANUT -- such as between $^{40}$Ca and $^{40}$Ar. This example is mainly a call to experimenters and model builders to pay attention to the requirements of this new field.

\subsection{Effect on the nuclear cascade}
Let us now discuss the case of strong disintegration: We discuss  the effect of photo-disintegration of $^{56}$Fe in a GRB for the TALYS model, where the parameters have been chosen such that the target photon density is high enough to observe the cascade -- see figure caption for the chosen shell parameters. The GRB model methodology closely follows \Refs~\cite{Hummer:2011ms,Baerwald:2013pu}, extended by solving the coupled partial differential equation system for all considered nuclear species. The considered hadronic energy losses include photo-disintegration, photo-meson production, beta decay, spontaneous emission of nucleons, and pair production, in addition to synchrotron and adiabatic losses. Initially 481 isotopes with 41000 inclusive disintegration channels are considered, and recursive automatic reduction techniques are used to identify the dominant contributions; see the Methods section for details. While we focus in this study on the photo-disintegration regime (and we have chosen the minimal photon energy such that photo-disintegration will always dominate), the photo-meson production is based on a superposition model based on 
SOPHIA~\cite{Mucke:1999yb}, which is state-of-the-art in the literature~\cite{Anchordoqui:2007tn,Murase:2008mr,Kampert:2012fi}. Note, however, that this model cannot capture multi-nucleon effects, which may turn out to be important if photo-disintegration is dominated by the photo-meson regime at the highest energies; such a scenario can be invoked if there is a minimal photon energy cutoff, which may come from synchrotron self-absorption~\cite{Murase:2008mr}. We show in \figu{nuclei} the results for the PSB and TALYS models. For the PSB model there is only one isotope for each $A$; neutrons and protons are strongly populated, since they are created in each interaction. The TALYS model predicts the population of isotopes off the main diagonal, and in addition, significantly more light (such as $^4$He) and intermediate ($10 \lesssim A \lesssim 20$) nuclei than the PSB model. We also show the TALYS model for different systematics assumptions: the ``systematic offset'' and ``random \revise{offsets}'' cases \revise{(the random offset example depends on the realization, and here one such realization is shown; the cross sections are randomly picked for each isotope separately)}. The  uncertainties in the cross sections are inspired by \figu{dis40}, which are of the order of a factor two \revise{in the cross sections near the peak of the GDR. The lower and higher energy peaks are not present in any of the models and can be considered as artifacts of the measurements. By exploring the information in the EXFOR database, a conservative choice is then to} vary all unmeasured absorption cross sections randomly between $0$ and $2$ of their nominal values, and all absorption cross sections with only some partial information (inclusive channels) between $0.5$ and $1.5$. Note that we re-scale all absorption and inclusive cross sections for one isotope with the same number, as we obtain otherwise inconsistencies between the escape and re-injection rates. We can read off from the figure that certain isotopes will be more highlighted than others, while the total cascade looks similar to the nominal TALYS model. For the ``systematic offset'' panel, we systematically choose the minimal values of the ranges above in the un- or partially measured cases, which leads to a systematic suppression of the nuclear cascade. Here the nuclear cascade will stop because of the cross section suppression, and intermediate and light elements will not be populated.
The population of isotopes in the cascade development shown in \figu{nuclei} is related to the example of $^{56}$Fe injection. 
By considering a different injected nucleus, the isotope chart will be differently populated: if the cross section of injected nuclei is not well described, a systematic error will propagate into secondary particle spectra.

\begin{figure}[bt]
\begin{center}
 \includegraphics[width=0.45\columnwidth]{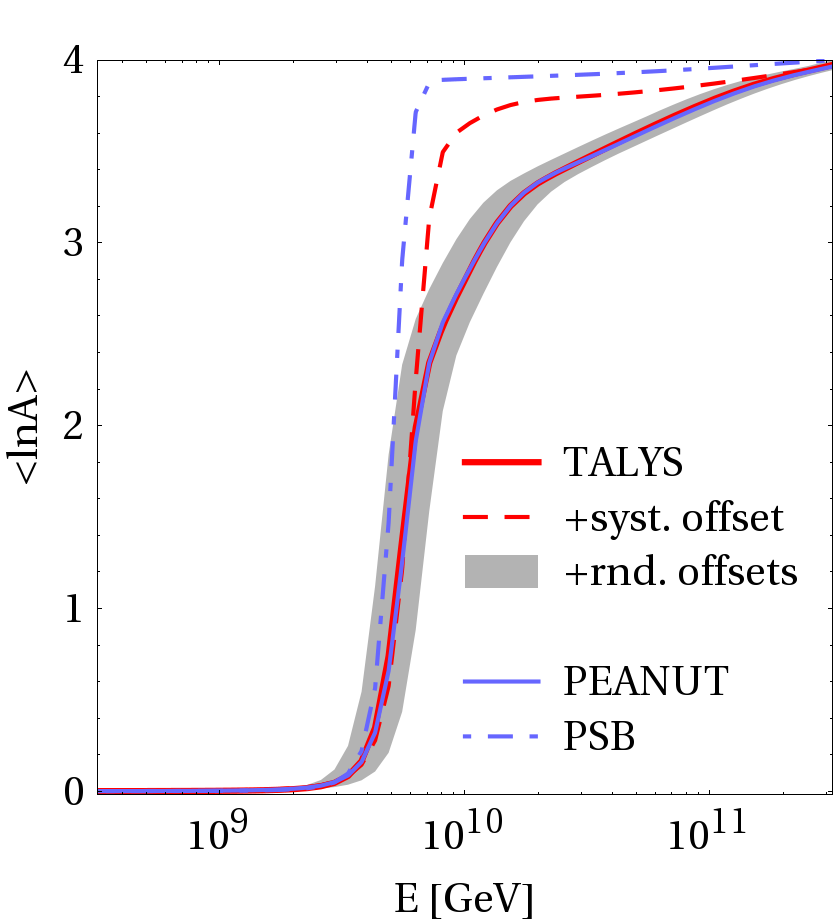} 
\end{center}
\caption{\label{fig:comp} Emitted cosmic ray composition from a GRB  as a function of observed energy without propagation effects for various nuclear disintegration models (same parameters as in \figu{nuclei},  iron injection only; see main text for details of the models). Here it is assumed that cosmic ray nuclei either escape as neutrons from the source (which leads to escape spectra $\propto E^{-2}$), or as charged particles if they can reach the outside of the source within their Larmor radius (``direct escape'', which leads to harder escape spectra $\propto E^{-1}$, see \Ref~\cite{Baerwald:2013pu})\revise{, that is consistent with what found in \cite{Aab:2016zth}}. 
}
\end{figure}

It is interesting to discuss the consequences for physical observables; we show the emitted cosmic ray composition from the GRB in \figu{comp}. First of all, it is noteworthy that the observed UHECR composition~\cite{Porcelli:2015} can be roughly reproduced from a single GRB shell and iron injection only, and will be even smoother after propagation. More sophisticated interaction models (TALYS, PEANUT) produce similar results, and even random cross section offsets within the model uncertainties do not qualitatively affect this picture, as the random variations seem to average out over the large number of involved isotopes (see shaded region for 100 ensemble models). However, the lack of/feeddown into light and intermediate elements produced in the PSB and ``systematic offset''  (systematic suppression of all unmeasured cross sections) models leads to strong deviations at about $10^{10} \, \mathrm{GeV}$ from the baseline models. \revise{For example, compare the PSB and TALYS models for $N \lesssim 17$: there are still many green-filled (compared to yellow-filled) isotopes in the TALYS model, and especially the low-$A$ isotopes ($^4$He and lighter) are enhanced in the TALYS model. This occuption of the intermediate mass isotopes leads to a smoother transition of the ejected composition for the TALYS model given that the ejected spectra of the charged cosmic rays are harder than $E^{-2}$ (see caption of \figu{comp} and Methods section for details on the cosmic ray escape, which changes the spectrum from injection to ejection).}

From this example, we conclude that the interaction model needs to be more sophisticated than the PSB model with a prediction power good enough to avoid systematic offsets. Note that such systematic offsets can also occur if a subset of isotope cross sections in the model is adjusted to measurements, while leaving the bulk of other unmeasured isotopes unmodified. Since neutrons can easier escape from the astrophysical source than protons, the cosmic ray composition is sensitive to an artificial imbalance between proton- and neutron-rich elements in the cascade. In addition, note that we expect other observables to be sensitive to the disintegration model as well, such as the neutrino production off nucleons versus heavier nuclei. The photo-meson production can, depending on the radiation densities and interaction model, be either dominated by the primary nuclei, or split-off nucleons. Since photo-meson production off nucleons is better understood than off nuclei, more robust predictions on the neutrino flux can be obtained in these cases.

We derive a complete list of isotopes for which the absorption and inclusive cross sections are needed to cover  any cosmic ray astrophysics scenario involving injection elements with masses up to iron, that we separate in ``Astro, priority 1'' and ``Astro, priority 2'' in \figu{exfor_chart}; it is apparently consistent with the isotopes populated in \figu{nuclei}. These isotopes have been obtained by recursively following all possible paths from all possible injection elements (\revise{we inject the most abundant stable isotope for each $Z$}) with certain threshold multiplicities (0.2 for priority~1 and 0.05 for priority~2), and by repeating this procedure for different interaction energies (see also the Methods section). That cross section information should ideally come from measurements. Until then,  predictive models may be needed, complemented with or derived from measurements of absorption cross sections and leading branching ratios of stable and unstable nuclei. The comparison of the absorption cross section among different isobars  may be an interesting cross check for the models. These measurements could be obtained at gamma-ray facilities such as ELI-NP \cite{Filipescu:2015hfj}, newSUBARU \cite{newSUBARU} or HI$\gamma$S \cite{HIGS}.

\section{Discussion and conclusions}

Since the evidence for a heavy composition of the observed cosmic rays is condensing, understanding the origin of cosmic rays requires the description of astrophysical sources with high radiation densities in the presence of nuclei. The long-term vision to determine the injected cosmic ray composition \revise{in the sources} and to identify unknown astrophysical parameters by multi-messenger astronomy requires combined source-propagation models for cosmic ray nuclei. 

As a first step into that direction, we have identified the requirements from nuclear physics for the leading process governing the nuclear cascade both within the sources and during cosmic ray propagation -- which correspond to very different (non-thermal versus thermal) radiation fields. Our working hypothesis has been that both sources and propagation need to be described by a comparable level of complexity for these applications, and we have therefore developed \revise{new methods} capabable to describe the radiation processes in the sources including the full nuclear disintegration chain \revise{efficiently}.

As one key result, we have compared the situation of nuclear measurements (red/yellow in \figu{exfor_chart}) with the input needed for cosmic ray astrophysics (blue in \figu{exfor_chart}). 
We have demonstrated that the measurements on the participating nuclear isotopes are sparse, as unstable nuclei are produced in the photo-disintegration -- which can live relatively long at extremely high energies where their lifetimes are Lorentz-boosted.

Although sophisticated nuclear models exist, such as TALYS, their prediction power for the considered isotopes 
seems limited to within about a factor of two. While the impact on individual photo-disintegration rates can be large, we have found that random fluctuations tend to average out in the nuclear cascade unless there are systematic effects not probably accounted for, such as offsets between neutron- and proton-rich elements, along the main diagonal, or a missed feeddown into lighter isotopes. We therefore propose systematic measurements to improve the predictability of unmeasured cross sections.
A long term goal could be measurements of total and inclusive cross sections of the blue-marked isotopes in \figu{exfor_chart}, or corresponding predictive information from nuclear theory.

Finally, we have demonstrated that -- while simple nuclear disintegration models for the sources may be sufficient in some cases -- there are observables which require a treatment at the level of complexity of TALYS or FLUKA within the cosmic ray source. For example, the ejected cosmic ray composition from a GRB deviates in the energy range where intermediate mass nuclei are produced in these models up to a factor of two in $\langle \ln A \rangle$. Similar consequences are expected for other observables, such as the secondary neutrino production, which requires further study.

We conclude that nuclear cosmic ray astrophysics will emerge as a new discipline which faces a new level of complexity compared to proton-only radiation models for the sources. If we really want to understand the physics of the sources and close the argumentation chain from the cosmic ray injection \revise{within the source}, over particle acceleration and radiation, to cosmic ray propagation, we need better models especially on the source side. We  have presented a radiation model with an unprecedented level of complexity for the nuclear disintegration in a gamma-ray burst, where the developed methods can be used in combined source-propagation models in the future, or in studies of the particle acceleration itself. While the astrophysical parameters and environments are highly uncertain, one can use such models to address the reverse question, \ie, use multi-messenger astronomy to derive the injection composition in the sources and to study the source parameters.

\revise{The radiation density in most of the possible UHECR sources dominates over the matter density; for this reason, the photonuclear reactions are the responsible for energy losses and production of secondary particles. However, for the case of Galactic cosmic rays, interactions with the interstellar medium have to be considered during propagation in the Galaxy. A similarity with what we studied in the present work }can be found from the point of view of the importance of nuclear physics. Secondary nuclei and antiparticles are produced by spallation of primary cosmic rays and carry information about origin of the primaries and transport in the Galaxy \cite{Coste:2011jc}. In particular, the most relevant uncertainties in modeling the antiproton flux, that constitutes also one of the prime channels for indirect searches of Dark Matter, come from the absence of measurements of p-He reactions \cite{Giesen:2015ufa,Kappl:2015bqa}. As a consequence, new input from nuclear physics could be of crucial importance also in \revise{nuclei-nuclei/nucleon processes, mostly relevant for} Galactic cosmic ray astrophysics.\\

\section{Methods}

Our GRB model follows \Refs~\cite{Hummer:2011ms,Baerwald:2013pu}, extended by the explicit simulation of the nuclear cascade. Here we only sketch the key ingredients, whereas details of the computation method and the efficient treatment of the particle interactions are presented in \cite{Biehl:2017zlw}. 

We solve a system of partial differential equations, which, for particle species $i$ (in our case nuclei), reads
\begin{equation}
\frac{\partial N_i}{\partial t} = \frac{\partial}{\partial E} \left( - b(E) N_i(E) \right) - \frac{N_i(E)}{t_{\text{esc}}} + \tilde Q_{ji} (E) \, , \label{equ:master}
\end{equation}
where $b(E) = E t^{-1}_{\text{loss}}$ with the energy loss rate $t^{-1}_{\text{loss}}=E^{-1} |dE/dt|$ and $t^{-1}_{\text{esc}}$ the escape rate. The equation is solved for the differential particle densities $N_i$ as a function of time, evolved until the steady state is reached. The coupled system arises because the injection term 
\begin{equation}
 \tilde{Q}_{ji}(E)= Q_i(E) + Q_{ji}(E) \label{equ:inject}
\end{equation}
allows for injection from an acceleration zone $Q_i$, as well as for injection from other species $j$ with the term $Q_{ji}$, such as from photo-disintegration. In this work, we only inject $^{56}$Fe into the system with a power law $E^{-2} \, \exp\left(-(E/E_{\mathrm{Fe,max}})^2 \right)$, where $E_{\mathrm{Fe,max}}$ is determined by balancing the dominant energy loss process with the acceleration rate $t^{-1}_{\text{acc}}=\eta c/R_L$. Here $R_L$ is the Larmor radius and a relatively high $\eta \simeq 10$ has been chosen to reproduce the composition transition observed by Auger~\cite{Porcelli:2015}. We also notice that choosing a milder cutoff in the injection spectrum reduces the required $\eta$. The target photons are assumed to follow a broken power law with a break at 1~keV in the shock rest frame between the power law index $-1$ (low energy) and $-2$ (high energy). Note that we choose the minimal photon energy low enough such that photo-disintegration will always dominate at the highest energies. The baryonic loading is assumed to be ten, which means that there is ten times more energy injected into iron compared to electrons, and energy equipartition between magnetic field, electron and photon energy is assumed.

For the nuclear energy losses, we include synchrotron losses, adiabatic losses, pair production losses, and photonuclear losses, where we distinguish between photo-disintegration (focus of this work, dominated by the GDR) and photo-meson production (dominated by the $\Delta$-resonance and other processes) by $\epsilon_r$.
The photo-disintegration is implemented in different models, as discussed in the main text of this work.
 We include beta decays of radioactive isotopes and spontaneous emission channels which are relevant on the dynamical timescale~\cite{IsotopeData}. 
We furthermore include photo-meson production based on SOPHIA~\cite{Mucke:1999yb} using an improved and extended version of \Ref~\cite{Hummer:2010vx} with a superposition interaction model assuming that the cross section scales $\propto A$. While such an approach is state-of-the-art in the literature, see \eg\ \Refs~\cite{Anchordoqui:2007tn,Murase:2008mr,Kampert:2012fi}, it is clear that it needs improvement from the interaction model perspective. Note that photo-meson production leads to the disintegration of nuclei as well, where the characteristics (secondary nuclei multiplicities and distributions) can be different; however, for the chosen parameters, the photo-disintegration dominates. Especially a minimal photon energy spectrum cutoff can complicate this picture at the highest energies, which may come from synchrotron self-absorption~\cite{Murase:2008mr}, as the high-energy nuclei will not find interaction partners (photons) anymore at the GDR relevant energy. In this case, the photo-disintegration will be dominated by the photo-meson energy regime. For example, in our superposition model, it is implied that the interactions of neutrons and protons are point-like, which means that the interaction probabilities are proportional to the number of neutrons and protons, respectively, and that the interacting nucleon splits off the nucleus. For this assumption, isotopes further off the main diagonal can be populated. Since the underlying (nuclear) physics cannot be captured by the models discussed here and these assumptions have to be cross-checked against leading codes, a more detailed study is beyond the scope of this work.

In order to discuss the ejected cosmic ray composition, we assume that cosmic rays can either escape as neutrons (which are not magnetically confined) or, directly from the borders of the production region within their interaction length or Larmor radius (whatever is smaller)~\cite{Baerwald:2013pu}, which leads to relatively hard spectra acting as a high-pass filter~\cite{Globus:2014fka}. This is, in a way, the most conservative assumption one can make, as, for instance, diffusion would enhance the escape at lower energies, or the magnetic fields may decay quickly enough that all particles can escape. Note that assuming Bohm-like diffusion for a steady source leads to a similar prediction. \revise{These hard spectra are the ones that have to be considered as inputs for the propagation in the extragalactic space. This is for example in agreement with the conclusions of \cite{Aab:2016zth}, where a hard spectral index for the sources of cosmic rays is found as a result of a combined fit of the spectrum and composition of UHECRs.} More detailed discussions of the astrophysical implications including the neutrino production are presented in \cite{Biehl:2017zlw}.

For an efficient computation, we need to pre-select the isotopes which are dominantly populated. 
Our isotope selection scheme is a fully automated recursive algorithm starting (in this case) with $^{56}$Fe, following all disintegration and beta decay paths  recursively. We also allow for mixed paths, meaning, for instance, that a disintegration may be followed by beta decay and then disintegration again. Since one isotope has typically very many (disintegration) branchings into daughters, some of these being very small, we impose a cutoff on the secondary multiplicity (number of secondaries produced on average) at a value of 0.01. As a further complication, the multiplicities, \ie, the distribution of secondaries, depend on $\epsilon_r$ (photon energy in nucleus' rest frame). In order to be as inclusive as possible, we therefore repeat the procedure for several values of $\epsilon_r$ in the relevant interaction energy range (actually we integrate over the pitch angle between nucleon and photon and use the pitch angle averaged multiplicities; the pitch angle enters the relationship between the photon/nucleus' energies and $\epsilon_r$, see the Supplementary Material).
We also tested other methods (such as based on the fractional contribution or picking the leading channels) with similar results. However, the result will depend on the choice of the control parameter, which is the threshold multiplicity here. As it turns out that intermediate mass number isotopes have a large number of daughter nuclei with relatively small multiplicities off the main diagonal, one cannot pick a too large threshold value as it would stop the cascade. Fast spontaneous emitters (faster than any other process at any given energy) are automatically integrated out by recursively following their decay paths to the next isotope which is explicitly treated. For example, for the TALYS disintegration model, the software uses 481 isotopes, about 41000 inclusive disintegration channels and about 3000 photo-meson channels as primary input, selects 233 isotopes, and attaches 4943 disintegration, 10 beta decay (relevant on the GRB timescale) and 1344 photo-meson channels to them. With this level of complexity and the explicit treatment of the nuclear cascade, it is clear that our technology is currently more advanced that any other method in the literature for the simulation of the sources of the UHECR nuclei themselves. 

The choice of isotopes in \figu{exfor_chart} labeled ``Astro, priority 1'' and ``Astro, priority 2'' has been obtained with a similar method by repeating the above recursive procedure for a number of possible injection elements: we inject for each 
atomic number the most abundant stable isotope. The multiplicity threshold has been chosen to be 0.2 for priority~1 and 0.05 for priority~2. As these numbers are larger than the threshold above, this list of isotopes is smaller than the one actually used for the GRB computation. However, one can easily see by comparing the blue isotopes in \figu{exfor_chart} with the ones in \figu{nuclei} for the TALYS model, that these are actually the most populated ones in a realistic simulation. Note that the main difference between these plots is that the extension of the nuclear cascade to low mass numbers depends on the energy in the GRB, and that only iron has been injected in the GRB case. We expect that our isotope priority selection is robust in the sense that taking into account different $\epsilon_r$ and being inclusive in the selected isotopes, they hardly depend on the shape of the target photon spectrum.

Additional information about cross section terminology and computation of disintegration lengths can be found in the Supplementary Material.\\

{\bf Acknowledgments.}  We would like to thank Alfredo Ferrari for enlightening discussions and the permission 
to use PEANUT for our purpose. We thank also B. K\"ampfer, A.R. Junghans and R. Schwengner from Helmholtz-Zentrum Dresden-Rossendorf for interesting discussions.

This project has received funding from the European Research Council (ERC) under the European Union’s Horizon 2020 research
and innovation programme (Grant No. 646623).


\newpage

\begin{appendix}
\section{Cross section terminology and computation of disintegration lengths}
\label{sec:term}

Numerical solutions of our system of Boltzmann transport equations (see the Methods section in the main text) require knowledge about the rate of interactions and the average outcome of the bulk of all possible reactions. The interaction rate  $\Gamma_j$ of the primary nucleus $j$ is used to compute the secondary injection and primary escape flux. The interaction rate depends on the interaction partner, the photon, as
\begin{eqnarray}
\Gamma_j (E_j) & = & \int d \varepsilon \int\limits_{-1}^{+1} \frac{d \cos \theta_{j \gamma}}{2} \, (1- \cos \theta_{j \gamma}) n_\gamma(\varepsilon,\cos \theta_{j \gamma}) \, \sigma_j^\text{abs}(\epsilon_r) \, .
\label{equ:Agamma}
\end{eqnarray}
Here $n_\gamma(\varepsilon,\cos \theta_{j \gamma})$  is the photon density as a function of photon energy $\varepsilon$ and the (pitch) angle between the photon and proton momenta $\theta_{j \gamma}$, $\sigma^\text{abs}_j(\epsilon_r)$ is the absorption cross section and 
\begin{equation}
\epsilon_r=\frac{E_j \, \varepsilon}{m_j} (1 - \cos \theta_{j \gamma})
\label{equ:epsilon_r}
\end{equation} is the photon energy in the parent rest frame (PRF) or nucleus rest frame in the limit $\beta_A \approx1$, and can be directly related to the available center-of-mass energy. The interaction itself, and therefore $E_j$ and $\varepsilon$, is typically defined in the shock rest frame (SRF). \equ{epsilon_r} clearly shows that the relevant energy range in the PRF is defined by the product of the target photon's energy times the nucleus energy in the SRF and is therefore highly dependent on both, the choice of the target photon and the injection spectrum. In our models we assume isotropy $n_\gamma(\varepsilon,\cos \theta_{p \gamma}) \simeq n_\gamma(\varepsilon)$ of the target photon distribution. This means that, for GRBs, we compute the interactions in the shock rest frame where isotropy occurs if the target photons come from self-consistent internal radiation processes (such as  synchrotron emission of electrons/positrons). Isotropy smears out the substructure of the energy dependence of $\sigma_\text{abs}$.

The injection $j \rightarrow i$ of secondaries $Q_{ji}$, which we need in Eq.~2 (Methods section), is given by
\begin{equation}
Q_{ji}(E_i) = \int dE_j \, N_j(E_j) \, \frac{d \Gamma_{ji} (E_j)}{d E_i} \, ,
\label{equ:prod}
\end{equation}
where $d \Gamma_{ji} (E_j)/d E_i$ can be obtained from \equ{Agamma} by replacing the total absorption cross section with the differential inclusive cross section
\begin{equation}
 \sigma_j^\text{abs}(\epsilon_r) \rightarrow \frac{d \sigma^\text{incl}_{j \rightarrow i}}{dE_i} \equiv  \sigma^{\text{abs}}_j(\epsilon_r) \, \frac{d n_{j \rightarrow i}}{dE_i}  \, ,
\end{equation}
where $dn_{j \rightarrow i}/dE_i$ is the re-distribution function of the secondaries.
For the energy spectrum of secondaries, we make use of typical kinematics, where for example, for GRBs, the nuclei have PeV energies and the photons are in keV range. In that case the momentum transfer on the ejected secondaries is small compared to $E_j$, resulting in an approximate conservation of the relativistic boost. 
In that case, the re-distribution function can be approximated as a function of $\epsilon_r$ and $E_i/E_j$ as
\begin{equation}
\frac{d n_{j \rightarrow i}}{dE_i} \simeq  M_{j \rightarrow i}(\epsilon_r) \cdot \delta \left( E_i -  \frac{m_i}{m_j} E_j \right)  \, , 
\label{equ:nabsimple}
\end{equation}
where $M_{j \rightarrow i}$  is the average number of secondaries produced per interaction (also called ``multiplicity''). It can be related to the exclusive reaction cross sections, summed over all channels in which the species $i$ is produced, as
\begin{equation}
	M_{j \rightarrow i} = \frac{1}{\sigma^\text{abs}_j} \sum_{X_n} N_i \, \sigma^\text{excl}_j
 (\gamma,X_n + i)  \, ,
\end{equation}
where $N_i$ is the number of secondaries $i$ in that exclusive channel.
Note that the ratio between exclusive and total absorption cross section corresponds to the branching ratio.
 While for light fragments, $M$ can be larger than one, one expects for residual nuclei $M \lesssim 1$. 

Finally, note that fixed target experiments in photon beams typically measure the cross section in terms of PRF energy, and it is a frequently asked question up to which energy the cross section needs to be measured.
In astrophysical environments, a certain PRF energy can be related to the primary $E_j$ and photon $\varepsilon$ energies by the estimate $\epsilon_r \sim E_j \, \varepsilon/m_j$, neglecting the pitch angle averaging, see \equ{epsilon_r}. Astrophysical photon spectra typically exhibit features, such as a spectral break (such as for GRBs) or a maximum (such as for the CMB) at a certain $\hat \varepsilon$, whereas the cosmic ray primaries typically follow a power law. As a consequence, given a cross section $\sigma(\epsilon_r)$ at a certain energy $\epsilon_r$, primaries with the energy $E_j \simeq \epsilon_r m_j/\hat \varepsilon$ will be selected to interact, and the secondary spectra will exhibit spectral excesses corresponding to that primary energy. In that sense, $\sigma(\epsilon_r)$ as a function of the PRF is the required input  for astrophysical applications, and a definition of a target photon spectrum in the nucleus' rest frame does not make sense. Larger values of $\epsilon_r$ just mean that the spectral features will appear at higher energies in $E_j$ if the cross section is significantly large there. For all practical cases, one therefore needs the cross section in the region around the GDR up to the point where it significantly drops.
This can be seen \eg\ in \figu{dis23} by comparing the PEANUT cross section curve with the  measurement for $^{23}$Na, which are practically identical up to the maximally measured energy. The difference in the disintegration length is nevertheless large, as the contribution beyond $\epsilon_r\simeq$ 30~MeV, which is missing in the measurement case, cannot be neglected (and will be re-distributed by the pitch angle averaging). The direct correlation $E_j \propto \epsilon_r$,  translating cross section peaks into the disintegration rate, can be best seen in the CMB cases in \figu{dis23}.

\section{Situation on Experimental data and model calculations}
\label{sec:models}

Here we summarize the status on experimental data and theoretical models.

\subsection{Experimental data}

EXFOR \cite{Otuka2014272} is a nuclear reaction database, aimed to be a complete collection of experiments and theoretical evaluations performed during the last half century. The effort to maintain and standardize the data collection is a world-wide collaborative effort between various nuclear centers. In this work, we assume that this database is to a large extent complete and our result is not distorted by one or few missing entries.

The database contains experimental data points, model-evaluated data, ratios of cross sections and other data categories. For our purpose we apply the following criteria to the data selection:
\begin{itemize}
  \item only real experimental data, no evaluated cross sections;
  \item no reaction combinations, ratios or partial measurements; 
  \item full unfolded cross sections in barns;
  \item energy range of the measurement has to cover the GDR peak.
\end{itemize}

Further, we create two selections (EXFOR, $\sigma_{\rm abs}$) and (EXFOR, any), where the first includes only measured absorption cross sections $\sigma_{\rm abs}$ and the latter requires at least one inclusive cross section to be measured. With these criteria, EXFOR contains 14 absolute cross sections and 47 of the second category up to masses of $^{56}$Fe. (EXFOR, any) can be used to estimate the number of more certain absorption cross sections from nuclear reaction models. 

\subsection{Theoretical models}

Monte Carlo particle transport codes, such as FLUKA~\cite{Ferrari:2005zk,BOHLEN2014211}, MCNP \cite{MCNP_6} or PHITS \cite{phits} use evaluated nuclear data files to compute interaction rates and Intranuclear Cascade Pre-equilibrium Evaporation (ICPE) codes  for exclusive reaction cross/sections. FLUKA contains a comprehensive photo-nuclear cross section library based direct on data, photo-neutron evaluation and parameterizations for a large set of isotopes \cite{FLUKA_pd1,FLUKA_pd2}. It is coupled to the ICPE code PEANUT to generate final state distributions. Comparisons of PEANUT cross sections reveal a very good agreement to available data on $\sigma_\text{abs}$. MCNP employs in a similar manner ENDF-B-VII.1 \cite{Chadwick20112887} as evaluated nuclear data library, based on reaction model calculations using the GNASH code system. For photo-nuclear data file for PHITS is JENDL/PD-2004 \cite{JENDL-PD-2004}, which has often good agreement with data but also shows some interpolation or fitting artifacts for light and intermediate nuclei.

Our current baseline model is based on cross sections extracted from TALYS 1.8 for nuclei with $A \geq 12$ and CRPropa2 lighter nuclei. TALYS 1.8 was configured to use the Kopecki-Uhl generalized Lorentzian description, as it is recommended in \cite{Khan:2004nd} in order to better predict cross sections not only in the GDR region. Photo-disintegration cross sections for light nuclei in CRPropa2 \cite{Kampert:2012fi} are compiled from various references as follows: ${}^{9}$Be, ${}^{4}$He, ${}^{3}$He, ${}^{3}$H, and ${}^{2}$H as given in \cite{Rachen:1996ph}; ${}^{8}$Li, ${}^{9}$Li, ${}^{7}$Be, ${}^{10}$Be, ${}^{11}$Be, ${}^{8}$B, ${}^{10}$B, ${}^{11}$B, ${}^{9}$C, ${}^{10}$C and ${}^{11}$C as given in \cite{Kossov:2002}; ${}^{7}$Li as given in \cite{Varlamov:1986,Kulchitskii:1963}. In total we have from CRPropa2 19 primary isotopes and 69 inclusive cross sections. The TALYS 1.6 cross sections are currently used in CRPropa3, to predict photo-disintegration products of all available exclusive channels: proton, neutron, ${}^{2}$H, ${}^{3}$H,  ${}^{3}$He, ${}^{4}$He and combinations thereof; the parameters used to model the giant dipole resonance have been adjusted to match the cross sections reported in \cite{Khan:2004nd}, as explained in \cite{Batista:2016yrx}. The {\it SimProp} code uses an approximation scheme to implement TALYS cross sections, as reported in \cite{Aloisio:2016tqp}.

The PSB model \cite{Puget:1976nz}, which is tailored for cosmic ray propagation problems, represents each isobar with the most abundant stable isotope, starting from ${}^{56}$Fe down to ${}^{2}$H, excluding the unstable masses $5 \leq A \leq 8$. This model approximates the cross sections for one- and two-nucleon emission with a Gaussian shape in the low energy range (2 $\leq \epsilon_r \leq$ 30 MeV), while the cross sections for multi-nucleon emission in the high energy range (30 $\leq \epsilon_r \leq$ 150 MeV) are constant. In {\it SimProp} the list of isobars matches \cite{Puget:1976nz}, but the corresponding elements are assigned from Table 1 in \cite{Stecker:1998ib} by choosing the isotope that corresponds to the lowest energy threshold for the emission of one proton.
For the purpose of the present work, we have slightly changed the criterion for choosing the element with respect to what has been done in {\it SimProp} or in the original PSB model, that allowed for example for $A \rightarrow A^{'}=A-1$ with an increasing number of protons with respect to the parent nucleus, by choosing descending elements in the chain corresponding to the mass in the original list.

\section{Disintegration rates for alternative elements}\label{sec.otherelements}

\begin{figure*}
\begin{center}
 \includegraphics[width=1.0\textwidth]{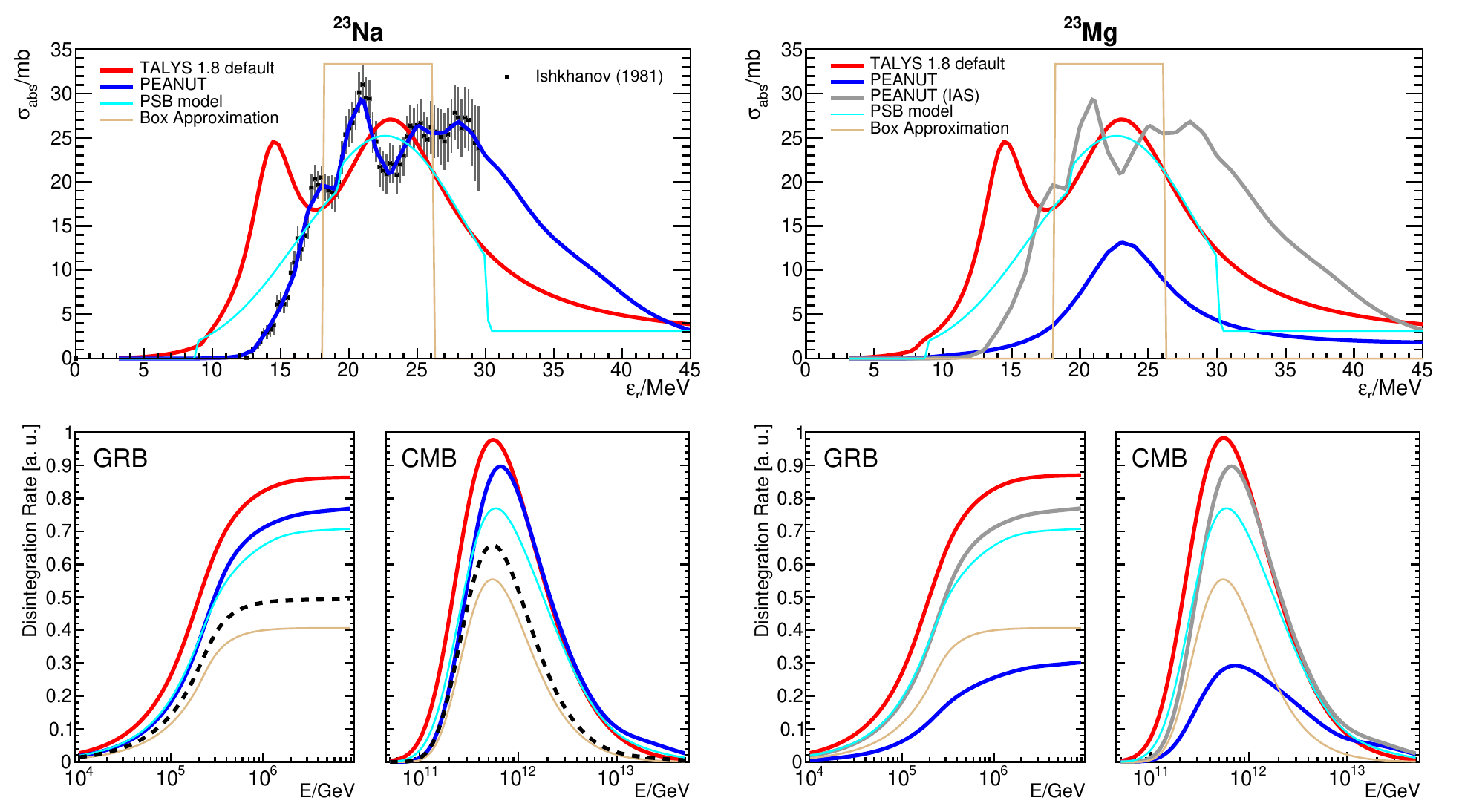} 
\end{center}
\caption{\label{fig:dis23} Same as in Fig.~2 (main text), for $A=23$.}
\end{figure*}
To illustrate a second example in addition to Fig.~2 (main text), we compare cross sections of ${}^{23}$Na and ${}^{23}$Mg in \figu{dis23}. Contrary to the double-magic ${}^{40}$Ca, ${}^{23}$Na is not expected to be spherical, where GDR cross sections can be sufficiently approximated using a single peak. Therefore, the models with only one main Lorentzian or Gaussian peak do not reproduce the measurement, as it is the case for the PSB model. Without dedicated Lorentzian fits, TALYS doesn't reproduce the shape of the GDR peak, adding an additional, unphysical, peak at $\sim 14$ MeV. The latter can be attributed to the strength parameter setting and can be potentially corrected through individual tuning. However, such modification would alter the model for a single isotope without improving the predictive power for other isobars. The PEANUT prediction reproduces very well the available measurement~\cite{Ishkhanov:1981} on ${}^{23}$Na, while the box model is insufficient and yields factor of two smaller disintegration rates. For the unknown ${}^{23}$Mg cross section, TALYS, PSB and the box model predict identical cross sections, while PEANUT falls back to a parametrized form, producing a significantly different result. Updated versions of PEANUT and FLUKA\footnote{Private communication with Alfredo Ferrari (07/2016)} estimate unknown cross sections with the Isobaric Analog State (IAS) approach, where nuclei with identical wave functions, and therefore, with a similar isospin and level structure are treated analogously. If such an equivalence case is detected, the model maps unknown cross sections of nuclei to a known IAS nucleus. The effect of IAS is illustrated in the right panel \figu{dis23} and results in up to a factor three higher disintegration rates in the case of PEANUT. If IAS is a valid concept for photon projectiles, several additional cross sections can become known to higher precision without the need for individual measurements.

\end{appendix}
\end{document}